\documentclass[reprint,superscriptaddress,amsmath,amssymb, nofootinbib, aps,twocolumn,pra]{revtex4-1}
\usepackage{graphicx}
\usepackage{multirow}
\usepackage{amsfonts}
\usepackage{bbold} 
\usepackage{verbatim} 
\usepackage{ulem} 
\usepackage{subfigure}
\usepackage[export]{adjustbox}
\usepackage{soul}
\usepackage{blindtext}
\usepackage[usenames, dvipsnames]{xcolor}
\usepackage{xifthen}

\newcommand{\bra}[1]{\langle #1 |}
\newcommand{\ket}[1]{| #1 \rangle}

\newcommand{\arxiv}[2][]{\ifthenelse{\isempty{#1}}{\href{http://arxiv.org/abs/#2}{{\tt arXiv:\allowbreak{}#2}}} {\href{http://arxiv.org/abs/#2}{{\tt arXiv:\allowbreak{}#2 [#1]}}}}

\begin{document}

\title{Quantum coherence and the principle of microscopic reversibility}


\author{K. Khan\smallskip}
\affiliation{Instituto de Física, Universidade Federal do Rio de Janeiro,
 Rio de Janeiro, RJ 21941-972, Brazil}
\author{W. F. Magalhães\smallskip}
\affiliation{Departamento de F\'{\i}sica, Universidade Federal da Para\'{\i}ba, 58051-900 Jo\~ao Pessoa, PB, Brazil}
\author{J. S. Araújo\smallskip}
\affiliation{Instituto de Física, Universidade Federal do Rio de Janeiro,
 Rio de Janeiro, RJ 21941-972, Brazil}
\author{B. de Lima Bernardo\smallskip}\email{bertulio.fisica@gmail.com}
\affiliation{Departamento de F\'{\i}sica, Universidade Federal da Para\'{\i}ba, 58051-900 Jo\~ao Pessoa, PB, Brazil}
\author{G. H. Aguilar\smallskip}\email{gabo@if.ufrj.br}
\affiliation{Instituto de Física, Universidade Federal do Rio de Janeiro,
 Rio de Janeiro, RJ 21941-972, Brazil}




\begin{abstract}
\noindent The principle of microscopic reversibility is a fundamental element in the formulation of fluctuation relations and the Onsager reciprocal relations. As such, a clear description of whether and how this principle is adapted to the quantum mechanical scenario might be essential to a better understanding of nonequilibrium quantum processes. Here, we propose a quantum generalization of this principle, which highlights the role played by coherence in the symmetry relations involving the probability of observing a quantum transition and that of the corresponding time reversed process. We study the implications of our findings in the framework of a qubit system interacting with a thermal reservoir, and implement an optical experiment that simulates the dynamics. Our theoretical and experimental results show that the influence of coherence is more decisive at low temperatures and that the maximum departure from the classical case does not take place for maximally coherent states. Classical predictions are recovered in the appropriate limits. 
     \\

\end{abstract}

\maketitle

\section{Introduction}

Fluctuation theorems (FT) are known to describe many general aspects of nonequilibrium thermal processes, and to bridge the gap between the reversible properties of the fundamental laws of physics and the irreversible nature of the macroscopic world \cite{evans,jarz,crooks,jarz2}. Nevertheless, despite their importance in fundamental physics and wide range of applicability in the study of many-particle systems, FT are formulated based only on two elements: the assumption of the Gibbs canonical ensemble to represent thermal equilibrium systems and the principle of microscopic reversibility \cite{crooks2,campisi}. The first concept is extensively discussed in many statistical mechanics textbooks, whereas the second, which is less widespread, predicts a symmetry relation between the probability of observing a given trajectory of a system through phase space and that of observing the time-reversed trajectory \cite{groot,chandler}. The microreversibility principle also plays a central role in the derivation of the celebrated Onsager reciprocal relations \cite{onsager1,onsager2}.

With the rapidly growing field of quantum thermodynamics, there has been an increasing effort to better understand FT when quantum effects become relevant \cite{talkner,esposito, aberg,manzano,kwon,khan,aguilar}. In this regard, one natural strategy is to initially investigate whether and how the principle of microscopic reversibility is modified by this classical-to-quantum transition. Unlike the classical domain, in the quantum regime we cannot know the simultaneous position and momentum of a particle with certainty, which obscures the very notion of trajectory, and it is also possible the formation of nonclassical correlations between system and environment upon interaction \cite{deffner,binder}. 
However, one can alternatively define stochastic trajectories of a quantum state in the Hilbert space to represent the dynamics of an open quantum system, whose interactions with the environment are conceptualized as generalized measurements \cite{breuer}. Still, depending on the commutation relation between the operators of the Hamiltonian of the system and those of the observables measured in an experiment, quantum coherence in the energy eigenbasis has to be considered, which prevents the system from having a well-defined energy \cite{korz,francica,santos,bert}. In a recent study of the interaction between coherent and thermal states of light in a beam-splitter, Bellini {\it et al.} verified the influence of quantum effects on the microreversibility condition, mainly in the low-temperature limit \cite{bellini}. 

Apart from being of central importance to the study of FT in the presence of quantum effects, so far very few works have addressed the microreversibility condition from a quantum-mechanical point of view \cite{bellini,agarwal1973open,nakamura2011fluctuation,tolman1925principle}. In this work, we propose a quantum generalization of the principle of microscopic reversibility. We consider the backward process of the system as resulting from the inverse unitary protocol applied to both system and reservoir, which connects the time-reversed states of the final and  initial states of the forward process. To test our model, we examine the dynamics of a two-level system coupled to a thermal reservoir at finite temperature, paying special attention to the mechanism in which the coherence influences the symmetry relation between forward and backward quantum trajectories. We also use an optical setup to simulate the open quantum system dynamics. Our setup allows the preparation of the system in an arbitrary qubit state  and the realization of projective measurements onto states with coherence. The experimental data agree with our theoretical predictions, which  confirm that the influence of coherence is more prominent at low temperatures, but in a nontrivial way. Our results recover the classical predictions in the appropriate limit.

Our paper is structured as follows. We begin in Sec.~II by briefly reviewing the principle of microscopic reversibility as applied to the simple case of a classical gas. In Sec.~III, we present our quantum-mechanical approach to the principle, and employ it to study the thermalization process of a qubit system interacting with a finite temperature environment. In Sec.~IV, we use our results to evidence the effect of coherence in the departure of our proposed quantum microreversibility relation from the classical picture. Sec.~V describes our experimental setup, in which we simulate the qubit thermalization example and investigate the role played by coherence in the quantum microscopic reversibility condition. We conclude in Sec.~VI with a summary of our theoretical and experimental findings.

\section{Principle of microscopic reversibility}
To begin, lets consider the rather simple but important scenario where the system is a classical gas in a container in contact with a thermal reservoir, in which some time-dependent parameter $\lambda(t)$ is controlled. In this case, the system can exchange heat $Q$ with the reservoir, and the manipulation of the control parameter may result in work $W$ being supplied to the system. To better illustrate this picture, let us assume a gas confined in a container with diathermic walls through which heat can be exchanged with the surroundings, and the work parameter $\lambda$ can be considered as the position of a movable piston, as depicted in Fig.~1. We define $z(t)$ as the time-dependent state of the system, which in our example is represented by a vector that determines all the instantaneous positions and momenta of the gas particles. Therefore, given the initial state $z_i$ and a protocol $\lambda(t)$, the dynamics of the system is completely specified by a trajectory $z(t)$ in the phase space, leading to a final state $z_f$.

\begin{figure}[ht!]
\includegraphics[scale=0.22]{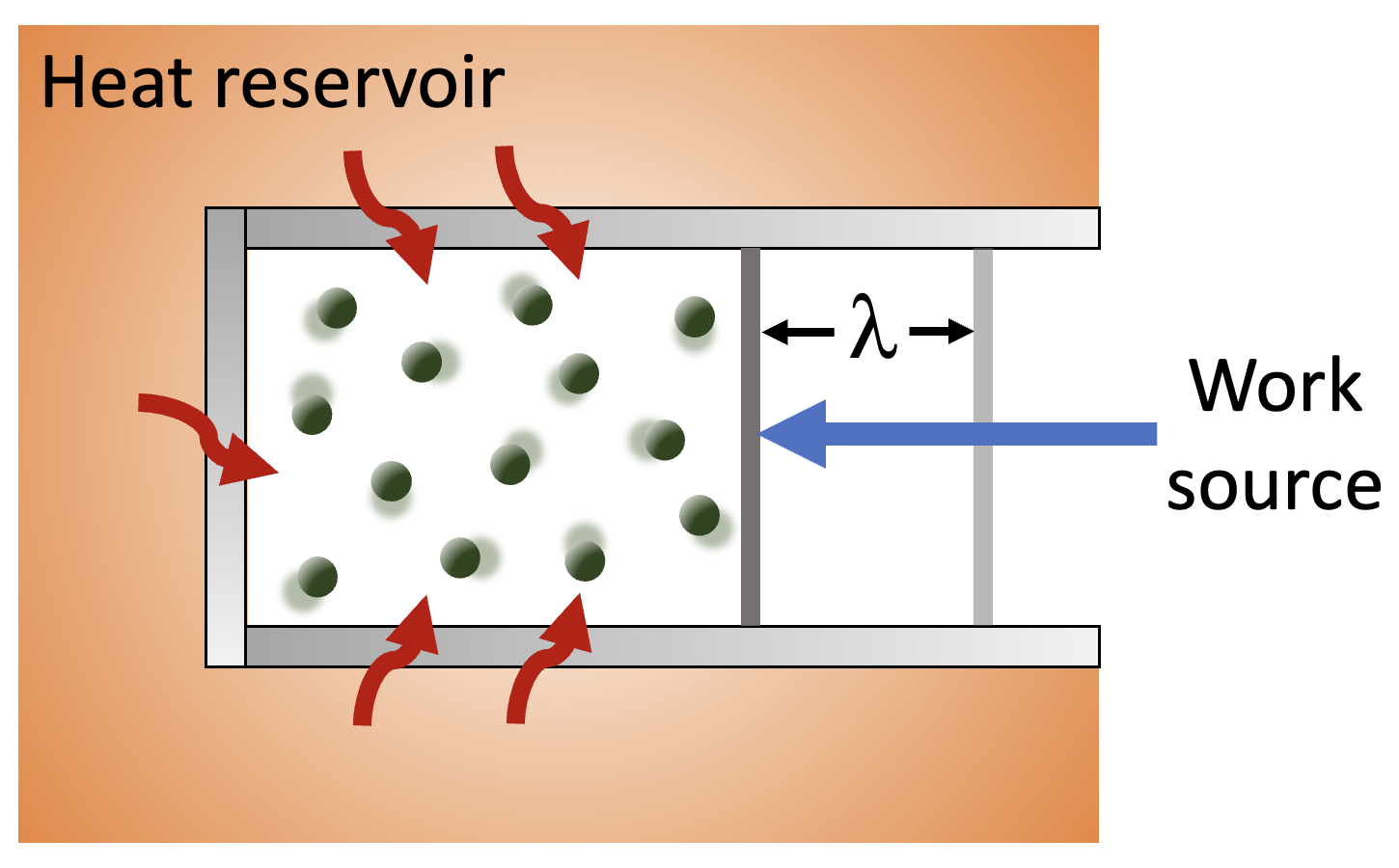}
\centering
\caption{Gas particles inside a container representing a general thermodynamic system. Transfer of heat between the gas and the reservoir is allowed through the boundary walls, and a work source acts by manipulating the position $\lambda$ of the movable piston.}
\end{figure}

On the other hand, we can also define a reverse trajectory $\Bar{z}(t)$, which can be visualized as a movie of the forward trajectory $z(t)$ played backwards. In this case, the process starts with the system in the state $\Bar{z}_f$ and ends in the state $\Bar{z}_i$. Here, an arbitrary state $\Bar{z}$ is produced by keeping the position of all particles of the state $z$ unchanged, but reversing the sign of all the momenta. If we assume that in the forward trajectory the system starts in the state $z_i$ at time $0$ and attains $z_f$ at time $\tau$, the reverse protocol of the work parameter satisfies $\Bar{\lambda}(\tau - t) = \lambda(t)$. We also consider that the heat reservoir is in thermal equilibrium during the entire process $z(t)$, so that it has the same influence on the system, independent of being in a forward or backward dynamics. With these definitions, the principle of microscopic reversibility states that the ratio of the probability of the forward trajectory from $z_i$ to $z_f$ to the probability of the corresponding backward trajectory from $\Bar{z}_f$ to $\Bar{z}_i$ is given by \cite{crooks,crooks2,groot}:      
\begin{equation}
\label{CMRP}
\frac{P_F(z_f|z_i)}{P_B(\Bar{z}_i | \Bar{z}_f)} = e^{-\beta Q},   
\end{equation}
where $Q$ is the heat absorbed by the system from the surroundings during the forward path, and $\beta = 1/(k_B T)$, with $k_B$ the Boltzmann constant and $T$ the temperature of the reservoir. This relation reveals that the probability of occurring a particular trajectory that generates dissipation ($Q < 0$) is exponentially greater than that of the corresponding reverse trajectory. If the control parameter is fixed and the gas is in thermal equilibrium with the reservoir, we have that $Q = 0$. 
In this case Eq.~(\ref{CMRP}) tells us that the probability of observing any given forward trajectory is equal to that of observing the corresponding backward trajectory.

\section{Quantum microscopic reversibility}

Based on the ideas presented in the previous section, we now investigate how the principle of microscopic reversibility can be extended to the realm of individual quantum trajectories with the control parameter kept constant, i.e., without the realization of work. At this level, some strictly quantum-mechanical conditions have to be considered. First, as already mentioned, the notion of quantum trajectory is not well defined \cite{sakurai}, but alternative approaches have been proposed, as for example the two-point measurement (TPM) protocol \cite{Talkner_2007}, which is the framework to be used here. 
Second, quantum measurements cause dynamical changes in the system that are never observed in classical objects \cite{jauch1964problem,muller2023six}. Third, depending on the commutation relation between the measured observables and the Hamiltonian of the system, the quantum dynamics could start or end in a state with coherence in the energy eigenbasis. In light of this freedom, the study of the influence of the initial and final coherence of the system on the microscopic reversibility relation is our main focus here.

Let us now present the quantum mechanical framework we use to study the microscopic reversibility. We consider a qubit system in contact with a large thermal reservoir composed of infinitely many thermal qubits at a given temperature. In this case, the system-reservoir dynamics can be considered as Markovian, and we assume that the system's evolution is described by the generalized amplitude damping channel (GADC) \cite{nielsen}. Physically, this model is a good approach for the study of qubit thermalization \cite{rosati,khatri}, and the Markovian to non-Markovian transition when increasing the reservoir size \cite{magalhaes}. The GADC has also been used to model a spin-1/2 system coupled to an interacting spin chain at finite temperature \cite{bose,goold}, thermal noise in superconducting-circuit-based quantum computing and in linear optical systems \cite{aguilar,chirolli}, and the effect of system-reservoir quantum correlations on thermodynamic quantities \cite{bert2}. In what follows we depict our perspective on what is a forward trajectory of a qubit submitted to the GADC, and the corresponding backward path. After this, we proceed to derive a quantum generalization of the microscopic reversibility and examine the role played by coherence.

\subsection{Forward process}

Let us describe the forward process in order to calculate the probability of observing it. We define the initial and final states of the two-level system as arbitrary pure qubit states:
\begin{equation}
    \label{ISS}
\ket{\psi_{k}}=\cos\bigg(\frac{\theta_{k}}{2}\bigg)\ket{g}+e^{i\phi_{k}}\sin\bigg(\frac{\theta_{k}}{2}\bigg)\ket{e}.
\end{equation}
The polar angle $\theta_k$ and azimuthal angle $\phi_k$ are used to locate the state $\ket{\psi_k}$ on the Bloch sphere, where the index $k=i,f$ indicates the initial and final states. The kets $\ket{g}$ and $\ket{e}$ represent the ground and excited states, which have energies $E_g$ and $E_e$, respectively. The density operators for the initial and final states are then given by $\hat{\rho}_{i}= \ket{\psi_{i}}\bra{\psi_{i}}$ and $\hat{\rho}_{f}= \ket{\psi_{f}}\bra{\psi_{f}}$.
As already mentioned, we consider a heat reservoir made by a large number of equivalent thermal qubits. These qubits are described by the state 
\begin{equation}
\label{ISE}
\hat{\rho}_{th} = w_{g}\ket{\mathcal{E}_g}\bra{\mathcal{E}_g}+w_{e}\ket{\mathcal{E}_e}\bra{\mathcal{E}_e},
\end{equation}
where $w_{j}=e^{-\beta E_{j}}/Z$, with the index $j=g,e$, such that $w_{g}+w_{e}=1$, and $Z = \sum_{j} e^{-\beta E_{j}}$ is the partition function. Note that the reservoir energy eigenstates, $\ket{\mathcal{E}_g}$ and $\ket{\mathcal{E}_e}$, also have $E_g$ and $E_e$ as the respective energies. This is necessary for the system and reservoir to interact through an energy-preserving unitary dynamics. We assume that system and environment are initially uncorrelated, such that the composite system starts out in the state $\hat{\rho}_{i}\otimes\hat{\rho}_{th}$.

We now pose the following question: What is the probability of the system starting out in the state $\ket{\psi_{i}}$ to evolve under the action of the GADC and then be measured at the end of the process in the state $\ket{\psi_{f}}$? The theory of open quantum systems tells us that this transition probability is given by:  
\begin{equation}
\label{PF}
P_{F}(\psi_{f}|\psi_{i}) = \mathrm{Tr}\bigg[\hat{U}\big(\hat{\rho}_{i}\otimes\hat{\rho}_{th}\big)\hat{U}^{\dagger}\big(\ket{\psi_{f}}\bra{\psi_{f}}\otimes\hat{I}_{\mathcal{R}}\big)\bigg],
\end{equation}
where $\mathrm{Tr}$ denotes the trace operation, $\hat{U}$ is the unitary operation that acts on both system and reservoir, which gives rise to the GADC acting on the system, and $\hat{I}_{\mathcal{R}}$ is the identity operator in the Hilbert space of the reservoir. Note that our selection of the initial and final states of the system is akin to that of the TPM scenario \cite{Talkner_2007}, where projective measurements are made both before and after the system's evolution. The unitary operation that supports the GADC is given by \cite{rosati,khatri,bert2}
\begin{equation}
\label{unitary}
\hat{U}=\begin{pmatrix}
1&0&0&0\\
0&\sqrt{1-p}&\sqrt{p}&0\\
0&-\sqrt{p}&\sqrt{1-p}&0\\
0&0&0&1\\
\end{pmatrix}.
\end{equation}
The damping parameter $p \in [0,1]$ represents the dissipation rate. Here, we are considering a Markovian dynamics, in the sense that the reservoir is made up of a large number of thermal qubits, each of which interacts only once with the system \cite{magalhaes}. In this case, we can assume $p = \mathrm{e}^{-t/\tau}$, where $t$ is the time and $\tau$ is a constant that characterizes the speed of the thermalization process \cite{fuji}.

With the above definitions, we are now able to calculate the right-hand side of Eq. (\ref{PF}), which provides
\begin{equation}
\label{PF2}
\begin{split}
P_{F}(\psi_{f}|\psi_{i}) =&\frac{1}{2
}\bigg\{\cos\theta_{f}\big[(1-p)\cos\theta_{i}+(1-2w_{e})p\big]\\
&+1+ \sqrt{1-p}\sin\theta_i\sin\theta_f\cos(\phi_{i}-\phi_{f})\bigg\}.
\end{split}
\end{equation}
One can also express this transition probability in terms of the quantum evolution time $t$ with the substitution $p = \mathrm{e}^{-t/\tau}$.
 
\subsection{Backward process}

Now we move on to describe the reverse  process and then proceed to calculate the corresponding backward transition probability. At this point, it is worth mentioning that there is not a unique definition for a reverse quantum process in the literature \cite{bellini,petz,barn,wilde,chib}. Here, we adopt the reverse process as a quantum dynamics whose initial state of the system is the time-reversed state obtained from $\ket{\psi_{f}}$, which evolves in contact with the thermal reservoir under the action of the reverse unitary operation $\hat{U}_{R} = \hat{U}^{-1} = \hat{U}^{\dagger}$. After this evolution, a measurement is performed such that we will be interested in the probability of obtaining the time-reversed state of $\ket{\psi_{i}}$. In this scenario, the transition probability for the reverse process is written as: 
\begin{eqnarray}
\label{PB}
P_{B}(\bar{\psi}_{i}|\bar{\psi}_{f}) =\mathrm{Tr}\bigg[\hat{U}_{R}\big(\hat{\rho}_{f}^{R}\otimes\hat{\rho}_{th}\big)\hat{U}^{\dagger}_{R}\big(\ket{\bar{\psi}_{i}}\bra{\bar{\psi}_{i}}\otimes\hat{I}_{\mathcal{R}}\big)\bigg],
\end{eqnarray}
where $\ket{\bar{\psi}}=\hat{\Theta}\ket{\psi}$ is the time-reversed state of $\ket{\psi}$, and $\hat{\Theta}$ the time-reversal operator \cite{sakurai}. Accordingly, we have
\begin{equation}
\label{TRstate}
\ket{\bar{\psi}_{k}}=\cos\bigg(\frac{\theta_{k}}{2}\bigg)\ket{g}+e^{-i\phi_{k}}\sin\bigg(\frac{\theta_{k}}{2}\bigg)\ket{e},
\end{equation}
from which we define the time-reversed  density operators $\hat{\rho}_{k}^{R}= \ket{\bar{\psi}_{k}}\bra{\bar{\psi}_{k}}$. In Eq.~(\ref{TRstate}) we implicitly assumed that the energy eigenstates $\ket{g}$ and $\ket{e}$ are invariant under time-reversal. Yet, we call attention to the fact that the initial system-reservoir configuration of the backward process is the uncorrelated state $\hat{\rho}_{f}^{R}\otimes\hat{\rho}_{th}$. This is the time-reversed state of $\ket{\psi_{f}}\bra{\psi_{f}} \otimes \hat{\rho}_{th}$, which is the final state of the forward process after the projective measurement. Note that the thermal state is invariant under time-reversal.

With these definitions, we can find that the transition probability for the reverse process is given by

\begin{equation}
\label{PB2}
\begin{split}
P_{B}(\bar{\psi}_{i}|\bar{\psi}_{f}) =&\frac{1}{2
}\bigg\{\cos\theta_{i}\big[(1-p)\cos\theta_{f}+(1-2w_{e})p\big]\\
&+1+ \sqrt{1-p}\sin\theta_i\sin\theta_f\cos(\phi_{i}-\phi_{f})\bigg\}.
\end{split}
\end{equation}
Again, the substitution $p = \mathrm{e}^{-t/\tau}$ allows us to express the transition probability as a function of the time $t$ elapsed from the beginning of the process.

\subsection{Symmetry relation between the forward and backward transition probabilities}

We can now use the information obtained from Eqs. (\ref{PF2}) and (\ref{PB2}) to derive the symmetry relation between the forward and backward transition probabilities. This is given by the following expression:
\begin{equation}
\label{PFPBratio}
\frac{P_{F}(\psi_{f}|\psi_{i})}{P_{B}(\bar{\psi}_{i}|\bar{\psi}_{f}) } =\frac{\cos\theta_{f}[(1-p)\cos\theta_{i}+(1-2w_{e})p]+\gamma}{\cos\theta_{i}[(1-p)\cos\theta_{f}+(1-2w_{e})p]+ \gamma},
\end{equation}
where $\gamma=1+\sqrt{1-p}\sin\theta_i\sin\theta_f\cos(\phi_{i}-\phi_{f})$. We can write $w_{e}=e^{-\beta \Delta E}/(1+e^{-\beta \Delta E})$, with $\Delta E = E_e - E_g$, as the reservoir excited state population. As such, Eq.~(\ref{PFPBratio}) is our quantum generalization of the microscopic reversibility condition for a qubit thermalization dynamics.

We first illustrate this finding with the important example in which quantum coherence in the energy eigenbasis is absent for both the initial and final states, i.e., the forward transition $\ket{g} \rightarrow \ket{e}$. For this process, we have that the system absorbs heat from the reservoir, and the parameters are given by $\theta_{i}=0$, $\theta_{f}=\pi$, and $\phi_{i}=\phi_{f}$. With this, we can find that Eq.~(\ref{PFPBratio}) reduces to
\begin{equation}
\label{PFPB1}
\frac{P_{F}(\psi_{f}|\psi_{i})}{P_{B}(\bar{\psi}_{f}|\bar{\psi}_{i}) }= e^{-\beta\Delta E},
\end{equation}
which holds independent of the value of $p$. Since no work is done by the reservoir on the system, we have that the total change in the internal energy of the system is due uniquely to the transfer of heat to or from the reservoir, namely, $Q=\Delta E$. This allows us to write
\begin{equation}
\label{PFPB12}
\frac{P_{F}(\psi_{f}|\psi_{i})}{P_{B}(\bar{\psi}_{f}|\bar{\psi}_{i})} = e^{-\beta Q}.
\end{equation}
This result tells us that when there is no coherence in the initial and final states of the quantum process, the classical symmetry relation between the forward and backward transition probabilities is recovered [see Eq.~(\ref{CMRP})]. 

\section{Microscopic reversibility in the presence of coherence}

As we have seen above, when the initial and final states of a quantum process have no coherence, our quantum generalization of the microscopic reversibility principle recovers the classical limit independently of $p$. Thus, we are left with four key parameters that determine how it deviates from the classical limit: the coordinates $\theta_i$, $\phi_i$, $\theta_f$ and $\phi_f$, which characterize the initial and final states of the system. However, if we are interested in investigating the influence of coherence on the microreversibility behavior, we have to express the right-hand side of Eq.~(\ref{PFPBratio}) in terms of the coherence of the initial and final states, $C_i$ and $C_f$. To do so, we quantify coherence with the $l_1$-norm of coherence \cite{baum,streltsov}. For a quantum state $\hat{\rho} = \sum_{i,j} \rho_{ij}\ket{i}\bra{j}$, the $l_1$-norm of coherence is given by the sum of the absolute values of all the off-diagonal entries, say $C(\hat{\rho}) = \sum_{i \neq j} |\rho_{ij}|$, which for the case of the pure qubit state given in Eq.~(\ref{ISS}) yields $C_{k}(\rho) = \sin \theta_{k}$, with $k=i,f$. Therefore, the quantum microreversibility relation of Eq.~(\ref{PFPBratio}) can be given in terms of the initial and final coherence of the system just by substituting $\cos \theta_k$ by $\sqrt{1-C^{2}_{k}}$ when $0 \leq \theta_k \leq \pi /2$, and by $-\sqrt{1-C^{2}_{k}}$ when $\pi /2 < \theta_k \leq \pi$. In this case, we also have $\gamma=1+\sqrt{1-p}C_i C_f\cos(\phi_{i}-\phi_{f})$. Moreover, the heat exchanged can be written in terms of the initial and final polar angles in the form $Q = \mathrm{Tr}[\hat{H}(\hat{\rho}_f - \hat{\rho}_i)] = \Delta E (\cos \theta_i - \cos \theta_f )/2$, which with the above relations can also be expressed as a function of the initial and final coherences. 

In order to better display our results, we define the following {\it deviation factor}: 
\begin{equation}
\label{Gamma}
\Gamma  = \left[ \frac{P_{F}(\psi_{f}|\psi_{i})}{P_{B}(\bar{\psi}_{f}|\bar{\psi}_{i})}\right] e^{\beta Q}.  
\end{equation}
When $\Gamma = 1$, the classical microscopic reversibility relation of Eq.~(\ref{CMRP}) holds. If $\Gamma >1$, we have that the backward process becomes less likely to happen when compared to the classical case. In turn, $\Gamma < 1$ means that quantum effects make the backward process be more likely.

In Fig. 2 we show the behavior of $\Gamma$ as a function of the coherences $C_i$ and $C_f$ for the case in which the system {\it releases} heat into the reservoir during the forward process, $Q < 0$. We assumed $\pi /2 \leq \theta_i \leq \pi$, $0 \leq \theta_f \leq \pi /2$, $\phi_i = \phi_f$ and $p=1/2$. It is clear that the influence of coherence becomes more and more relevant as temperature decreases. When coherence is absent, $C_i = C_f = 0$, the classical behavior emerges for all temperatures, which agrees with the result of Eq.~(\ref{PFPB12}). The values of $C_i$ and $C_f$ for which the {\it minimum} deviation factor $\Gamma_{min}$ occurs changes with temperature. For instance, for $\beta \Delta E = 1$ we obtain $\Gamma_{min} \approx 0.66$ when $C_i \approx 0.74$ and $C_f \approx 0.61$. In turn, for $\beta \Delta E = 2$ we find $\Gamma_{min} \approx 0.39$ when $C_i \approx 0.73$ and $C_f \approx 0.53$. We also remark that the classical behavior is suddenly recovered when we approach the maximum coherence point, $C_i = C_f = 1$ (also obtained with $\theta_i = \theta_f = \pi/2$), for all temperatures. The reason is that the maximum coherence point corresponds to the only point in the diagrams in which no net heat is exchanged between the system and the reservoir, $Q = 0$. We also observe that in this case the initial and final states of the process are the same. Hence, we do not expect any preferred direction involving the forward and backward trajectories. 

It is worth mentioning that  the classical limit, $\Gamma = 1$, is always attained in the high temperature limit, $\beta \Delta E \ll 1$, regardless of the values of $C_i$ and $C_f$. Let us discuss the physical meaning of this result. Classically, in this regime, the forward and the corresponding backward processes occur with almost the same probability, as can be seen from Eq.~(\ref{PFPB12}). On the other hand, in the quantum case studied here, the GADC acts by moving any initial state towards the center of the Bloch sphere, which represents the maximally mixed state. Projective measurements made on this state provides any pure qubit state with the same probability. Therefore, similar to the classical case, forward and backward trajectories connecting any pure states have practically the same probability to occur. This justifies $\Gamma =1$ in the high-temperature domain.

\begin{figure}[ht!]
\includegraphics[scale=0.26]{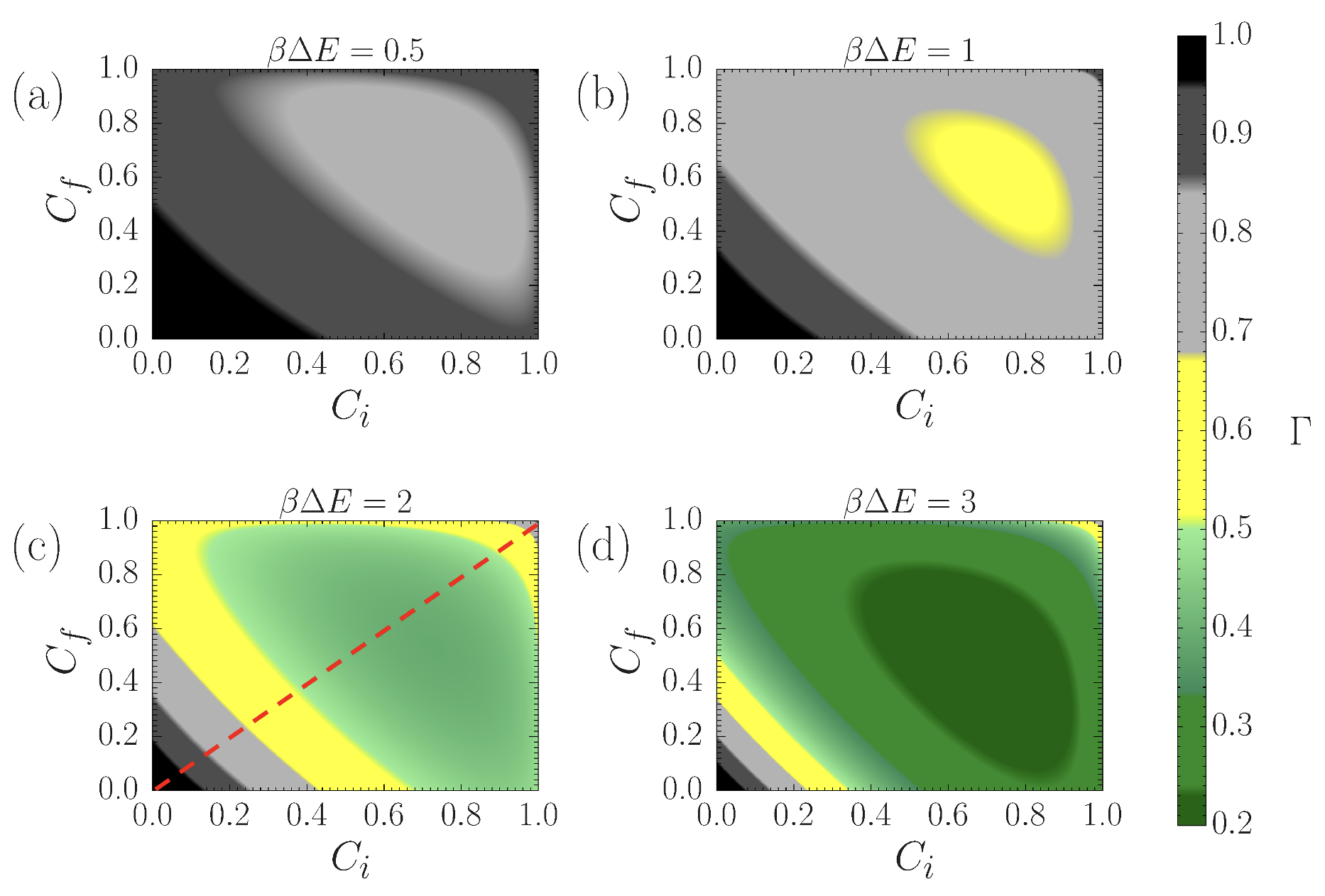}
\centering
\caption{Behavior of the deviation factor $\Gamma$ as a function of the initial and final coherences, $C_i$ and $C_f$, for the case in which the system releases heat to the reservoir in the forward process. (a) High-temperature scenario, where we show that $\Gamma$ deviates little from unit, indicating an approximately classical behavior. Diagrams (b), (c) and (d) show that the effect of coherence on the microreversibility principle becomes more prominent with lower temperatures. We assumed $\phi_i = \phi_f$ and $p=1/2$. The diagonal red cut in (c) is discussed in our experiment.}
\label{map1}
\end{figure}

\begin{figure}[ht!]
\includegraphics[scale=0.26]{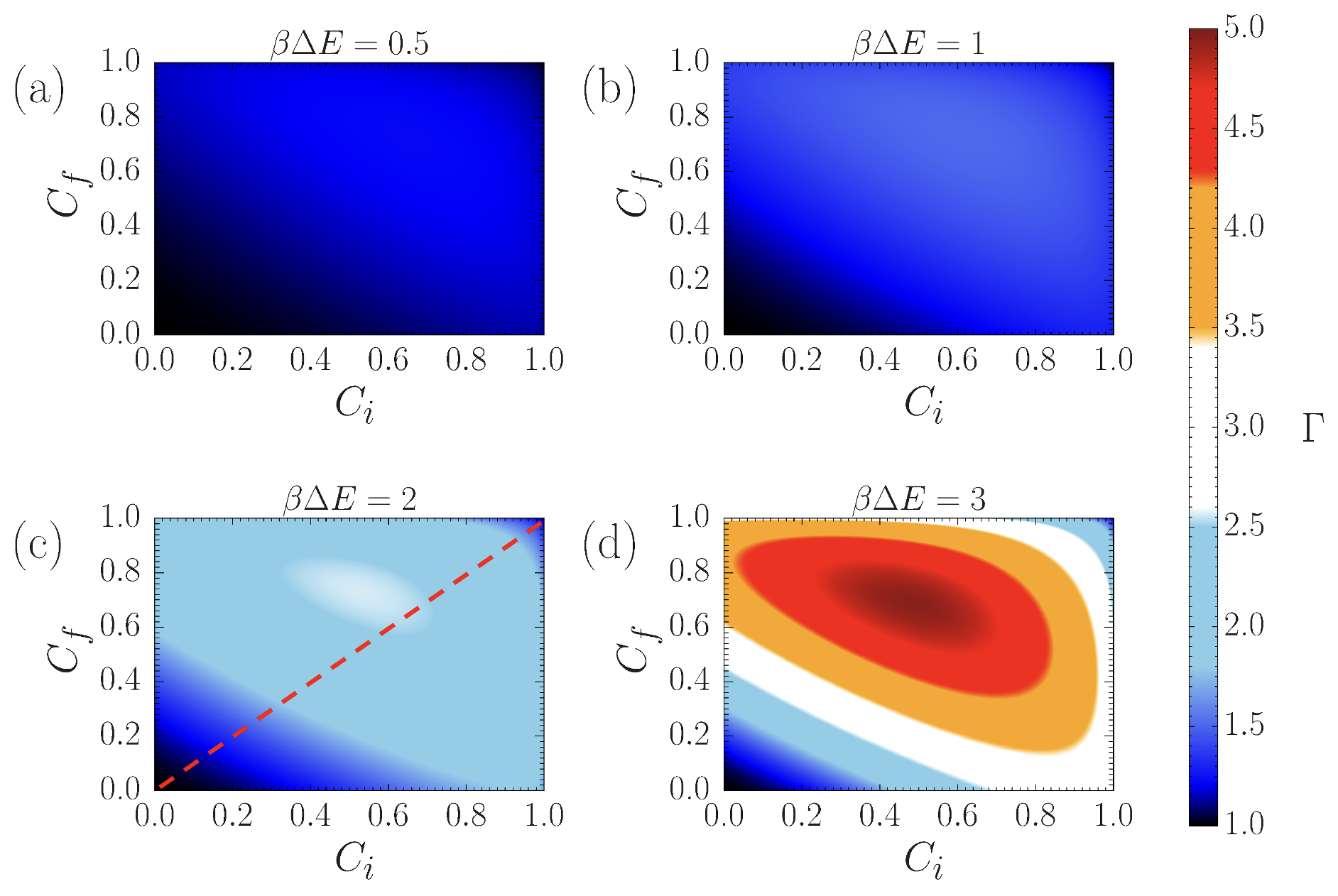}
\centering
\caption{Deviation factor $\Gamma$ as a function of $C_i$ and $C_f$ when the system absorbs heat from the reservoir in the forward process, considering $\phi_i = \phi_f$ and $p=1/2$.  (a) High-temperature regime showing that $\Gamma$ is only slightly bigger than one. Diagrams (b), (c) and (d) demonstrate  that, in the presence of coherences, the deviation from the classical microreversibility behavior becomes more relevant for lower temperatures. The diagonal red cut in (c) is discussed in our experiment.}
\label{map2}
\end{figure}

In Fig. 3, we show the behavior of $\Gamma$ as a function of $C_i$ and $C_f$ when the system {\it absorbs} heat from the reservoir in the forward process, $Q > 0$. Here, it was assumed that $0 \leq \theta_i \leq \pi/2$, $\pi /2 \leq \theta_f \leq \pi $, $\phi_i = \phi_f$ and $p=1/2$. In this case, we also observe that the classical microreversibility condition always emerges at high temperatures. The classical limit is also found to hold in the absence of coherence, $C_i = C_f = 0$, for all temperatures as predicted by Eq.~(\ref{PFPB12}). We call attention to the fact that the $C_i$ and $C_f$ values for which the deviation factor reaches a {\it maximum}, $\Gamma_{max}$, are temperature-dependent. For example, for $\beta \Delta E = 1$ we find $\Gamma_{max} \approx 1.51$ when $C_i \approx 0.61$ and $C_f \approx 0.74$, whereas for $\beta \Delta E = 2$ we get $\Gamma_{max} \approx 2.56$ when $C_i \approx 0.53$ and $C_f \approx 0.73$. The maximum coherence point, $C_i = C_f = 1$ ($\theta_i = \theta_f = \pi/2$), also manifests the classical behavior, $\Gamma = 1$, because in this case no net heat is exchanged with the reservoir, $Q = 0$. This result is valid for all reservoir temperatures. Here, we call attention to the fact that the ratio $P_{F}(\psi_{f}|\psi_{i})/P_{B}(\bar{\psi}_{i}|\bar{\psi}_{f})$ in Eq.~(\ref{PFPBratio}) is symmetric under the exchange between $\theta_i$ and $\theta_f$, with the result raised to the power of -1. This means that the panels in Figs.~\ref{map1} and~\ref{map2}, which correspond to same value of $\beta \Delta E$, are related by an inversion of the axis, followed by the transformation $\Gamma \rightarrow 1/\Gamma$. 

\section{Experiment}

\begin{figure*}
  \begin{minipage}{0.95\linewidth}
    \includegraphics[width=0.95\linewidth]{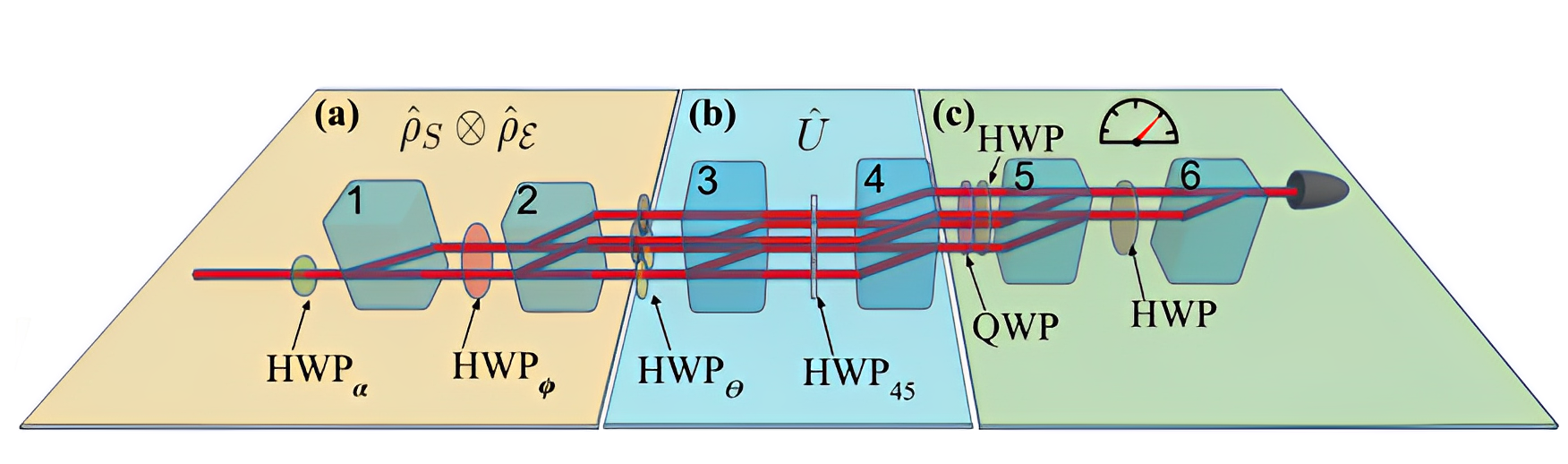}
     \caption{Aerial view of the two-layer interferometer. The first layer, representing the lower layer, contains the initial light propagation. The second layer is populated after BD2, whose output path modes codify the degrees of freedom of the systems and reservoir (see main text). Projective measurements are performed with BD5, BD6 and  wave-plates. The temperature of the reservoir is controlled by the angle $\phi$ while the damping parameter  is manipulated with the angle $\theta$. More details can be found in the Appendix A.} 
     \label{3DFIGURE}
  \end{minipage}
\end{figure*}


The  experimental setup to study microreversibility in the presence of coherences is depicted in Fig. \ref{3DFIGURE}. Here, the interaction between a qubit and a thermal reservoir is simulated using the GADC, which is implemented in a  two-layer optical interferometer. Optical simulations of open quantum systems have been important for the observation of effects, such as the sudden death of entanglement \cite{almeida2007environment}, sudden transition from quantum to classical decoherence \cite{mazzola2010sudden},  immunity  of correlations against certain decoherent processes \cite{xu2010experimental}, redistribution of entanglement between entangled systems and their local reservoirs \cite{aguilar2014experimental,farias2012observation}, among others \cite{liu2011experimental,chiuri2012linear,cuevas2019all,bernardes2015experimental,cialdi2017all,liu2018experimental,aguilar2015experimental}. All this analog simulations could be useful to bring the theory  closer to the physical systems and to provide valuable insights on how to observe the predicted effects in real-world scenarios.   Our setup can be divided into three main parts: a) the preparation of quantum states; b) the evolution of both the system and the reservoir through a unitary operation; and c) the execution of projective measurements. Detailed explanations of these three parts can be found in the Appendix A.

To experimentally assess quantum microreversibility, it is necessary to determine the probabilities of the forward and backward quantum processes. To obtain the probability of the backward process, the state of the photons is prepared now in the final state of the forward evolution $\ket{\psi_{f}}$ using the HWP$_{\alpha}$. Then, it interacts with the thermal reservoir by undergoing the  evolution influenced by the inverse unitary operation $\hat{U}^{-1}$, which is implemented with the two-layer interferometer. The backward process ends by performing a projective measurement onto the initial state of the forward process $\ket{\psi_{i}}$. This is accomplished by using the wave plates in the measurement stage. After computing  the probabilities   of both processes with Eqs.~(\ref{PF}) and~(\ref{PB}), our quantum version of the microscopic reversibility condition in the presence of coherences can be experimentally investigated.

\begin{figure}[ht!]
\includegraphics[scale=0.5]
{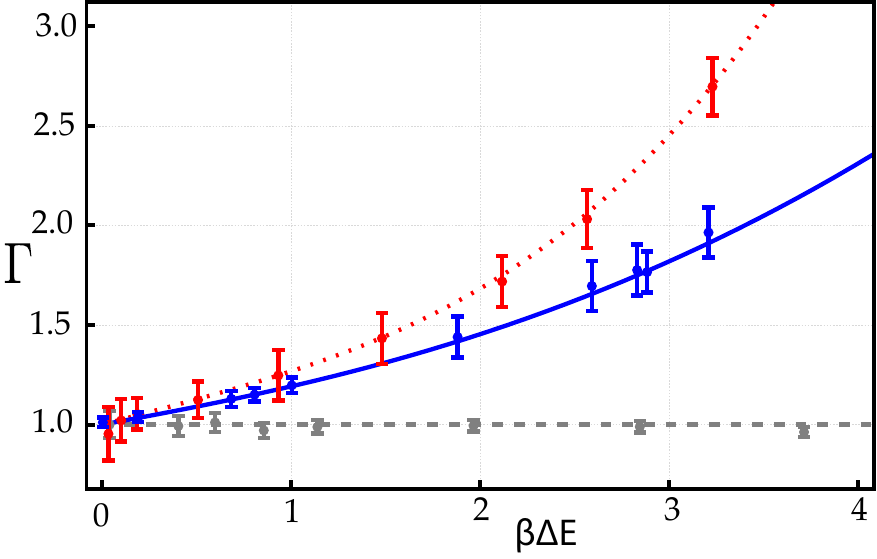}
\centering
\caption{Theoretical and experimental data of the deviation factor $\Gamma$ as a function of $\beta \Delta E$ for three different forward and backward transition cases. Gray dashed line and dots represent the classical  analog given by $(C_i,C_f)=(0,0)$. Red dotted line and points correspond to the transition from a maximally coherent state to a classical state, $(C_i,C_f)=(1,0)$. In blue, we have a representation of a transition from a maximally coherent state to a less coherent state,  $(C_i,C_f)=(1,0.87)$.} 
 \label{Experimental_results}
 \end{figure}

We started by studying the quantum condition for three different configurations of initial and final states as a function of  $\beta\Delta E$. The results  are shown in Fig. \ref{Experimental_results}. The data points, representing the experiment, exhibit a remarkable agreement with the  lines that are theoretical predictions. The error bars, calculated using a Monte Carlo simulation, indicate the uncertainty range within the experimental data. In the first case, corresponding to the gray dashed line and gray points, we showed the results for the system initiating in the state $\ket{g}$, evolving under the action of the GADC, and then being projected onto the state  $\ket{e}$ in the forward process. The reverse probability entails the system starting in the state $\ket{e}$ and, after evolution, being measured in the state $\ket{g}$. Remarkably, this case exemplifies a transition between two states with a classical analog, therefore, satisfying the classical microreversibility condition (dashed line). Moreover, for infinite  temperature  ($\beta= 0$),  the ratio $P_F/P_B$ is equal to one, indicating that there is no preferred direction between forward and backward process. This is also observed in the following two approaches.  

The second case, red points and dotted line, illustrates the probability of transitioning the system from a maximally coherent state to the excited state, specifically $1/\sqrt{2} (\ket{g}+\ket{e}) \longleftrightarrow \ket{e}$, and vice versa. 
 Here, we observe a significant deviation from the classical case, which shows the existence of quantum features in the microreversibility condition. Nevertheless, the maximum deviation from the classical case happens at low temperatures, as expected.  Blue points and solid line, representing the third case, describe a situation with coherence in the initial and final states, namely $1/\sqrt{2} (\ket{g}+\ket{e})$ and  $1/2( \ket{g}+ \sqrt{3} \ket{e})$, respectively. Again, there is a large deviation from the classical scenario.  Remarkably, in the latter two cases, this clear divergence from the classical microreversibility condition can be ascribed to the presence of coherence in both the initial and final states. It should be noted that in the first case the values of $C_i$ and $C_f$ are both equal to 0. The second case corresponds to $C_i = 1$ and $C_f = 0$, whereas in the third case $C_i$ and $C_f$ are 1 and approximately 0.87, respectively. It is surprising that the second case deviates more than the third one, considering that this latter possesses higher coherences. This shows that the deviation from the classical behavior does not necessarily increase as $C_i$ and $C_f$ increase. This can  also be observed in Fig.~\ref{map2}, where the second and third cases fall within the bottom right corner and in the upper part of the right edge, respectively, of each panel. For instance,   
in Fig.~\ref{map2}(d) one can observe that the second case is located in the white region, while the third case resides in the blue region, which corresponds to a lower deviation. Furthermore, the separation between the curves of cases 2 and 3 becomes progressively bigger as temperature decreases.

\begin{figure}[ht!]
\includegraphics[scale=0.85]
{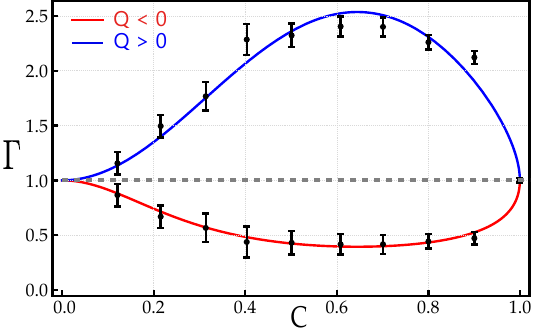}
\centering
\caption{Deviation factor $\Gamma$ as a function of the coherence $C$ when the system releases heat (red) and absorbs heat (blue). Experimental results are the black dots, and the theoretical solid lines come from the diagonal cuts in Figs.~\ref{map1}(c) and~\ref{map2}(c), which correspond to $\beta \Delta E=2$. The classical limit is represented by the dashed line.}
\label{Experimental_results2}
 \end{figure}


We also study the quantum microreversibility by varying the initial and final states while keeping the temperature constant. Throughout these experimental measurements, the system was prepared and measured using states described by Eq.~(\ref{ISS}). For each experimental run, we carefully adjusted the parameters $\theta_i$ and $\theta_f$ to ensure that the initial and final coherences coincide, $C_i = C_f = C$. To encompass the entire range of coherence values, we systematically varied $\theta_i$ from $\pi/2$ to $\pi$ and $\theta_f$ from 0 to $\pi/2$, thereby investigating the diagonal cut indicated in Fig.~\ref{map1}(c). The experimental results for these measurements are presented in Fig. \ref{Experimental_results2}. It is important to highlight that the result of the deviation factor $\Gamma$ for the heat absorption case (shown in red in Fig.~\ref{Experimental_results2}) can be obtained directly from the corresponding result of the heat release case (blue curve) with the mentioned symmetry properties of Eq.~(\ref{PFPBratio}). This illustrates that the heat absorption case essentially mirrors the heat release case. 
One can observe that these measurements also deviate from the  classical microreversibility condition ($\Gamma=1$). More precisely, our experimental results are below the classical value by over 17 standard deviations when the system is releasing heat, and more than 7 standard deviations when it absorbs heat. These results also indicate that the behavior of microreversibility in the presence of coherences is highly divergent from the classical counterpart. Moreover, according to our theory these deviations become even larger as temperature decreases.\\

\section{Conclusions and outlook}

In summary, we proposed a quantum-mechanical approach to the principle of microscopic reversibility with the aim to investigate how quantum effects impact the symmetry relation between the probabilities of observing a given process and that of the corresponding time reversed transformation. Based on this approach, we studied the microreversibility properties of the dynamics of a qubit system in contact with a heat reservoir at finite temperature, from which we identified the influence of the initial and final coherence. The system dynamics was modeled with the GADC, and the backward transformation was considered as resulting from the inverse of the unitary operation that acts on both the system and reservoir in generating the forward process. In particular, we have seen that, in all cases in which the forward process is such that the system {\it releases} heat to the reservoir, $Q<0$, the general effect of the presence of initial and final coherences in the energy eigenbasis of the system, $C_i$ and $C_f$, is to {\it increase} the probability of observing the backward process in comparison with the classical case. This fact was testified by the observation of a deviation factor, $0 < \Gamma < 1$. On the other hand, when the forward process is such that the system {\it absorbs} heat from the reservoir, $Q>0$, we verify that the presence of coherence contributes to {\it decrease} the probability of observing the backward process when compared to the classical case. This situation provides $\Gamma > 1$.

Our results also showed that the classical result of the microreversibility condition, Eq.~(\ref{CMRP}), is always recovered in three cases: i) when the initial and final state of the system in the quantum process have no coherence ($C_i = C_f =0 $), ii) in the high-temperature limit, and iii) when the system starts and ends the process in a maximally coherent state ($C_i = C_f = 1$). In our study, this last case corresponds to a situation in which the system does not exchange net heat with the reservoir, i.e., $Q=0$, which means that there is no preferred direction with respect to the forward and backward process. Another important finding of this work is that the effect of coherence becomes more significant in the departure of the quantum microreversibility condition from its classical counterpart as temperature decreases. In Ref.~\cite{bellini}, the authors reported a similar result by examining the behavior of coherent and thermal states of light mixed in a beam-splitter. Here, we observed that the points ($C_i,C_f$) at which the deviation factor is an extremum varies with temperature. Yet, in the limit of $\beta \Delta E \rightarrow \infty$, calculations have demonstrated that $\Gamma \rightarrow 0 $ in the $Q<0$ case of the forward process, and $\Gamma \rightarrow \infty $ in the $Q>0$ case. These results tell us that heat absorption processes in this regime are infinitely more likely to occur in the quantum scenario than in the corresponding classical situation. In fact, in the classical case the system is never allowed to absorb heat, so that reverse excitation processes are completely forbidden. Conversely, in the quantum case we have that the GADC reduces to the amplitude damping channel, in which the reservoir is considered to be at $T=0$ \cite{nielsen}. This channel acts by moving any initial state towards the state $\ket{g}$. Nevertheless, before attaining this state, there is always a non-zero probability of observing excitations events after the realization of the second projective measurement on the system.

We realized an experimental simulation of our findings with an all-optical setup. It consists of a two-layer interferometer, in which the path degrees of freedom encode the energy states of the system and reservoir, and the polarization degree of freedom acted as an ancillary qubit. For the preparation and evolution of the composite system-reservoir state, we utilize a set of wave-plates. The experimental results showed excellent agreement with our theoretical description and, in the $\beta \Delta E = 2$ case, data showed that the quantum microreversibility condition deviates from the classical counterpart by 7$\sigma$ (when the system absorbs heat) and 17$\sigma$ (when the system releases heat). Nevertheless, our theoretical results predict that this deviation can be larger by choosing lower temperatures or other cuts of the maps shown in Figs.~\ref{map1}(c) and~\ref{map2}(c). For example, in the $Q > 0$ case one could use a cut that passes through the maximum deviation point, which happens for $C_i \approx 0.61$ and $C_f \approx 0.74$, when $\beta\Delta E=2$.  Finally, with the microscopic reversibility assumption being an essential ingredient in the derivation of fluctuation theorems, we believe that the present results can provide an important insight in the understanding of the role of coherence in nonequilibrium quantum processes.

\vspace{1mm}
\acknowledgments
The authors acknowledge financial support from the Brazilian agencies Coordenação de Aperfeiçoamento de Pessoal de Nível Superior - CAPES (Finance Code 001), and Conselho Nacional de Desenvolvimento Científico e Tecnológico - CNPq (PQ Grants No.~310378/2020-6 and~307876/2022-5, and INCT-IQ 246569/2014-0). GHA acknowledges FAPERJ (JCNE E-26/201.355/2021) and FAPESP (Grant No. 2021/96774-4).

\appendix

\section{Experimental Configuration for the Study of Microreversibility in the Presence of Coherences}
In this appendix, we present a comprehensive explanation of the experimental setup employed to investigate microreversibility in the presence of coherences, as illustrated in Fig. 4 of the main text. 
This setup is categorized into three distinct parts. The initial part (a) is dedicated to the preparation of quantum states, while the second part (b) entails the evolution of both the system and the environment through a unitary operation. The final part (c) is responsible for the execution of projective measurements. The following text  provides the detailed explanation of each part.

\section*{a) ~ States preparation}
Starting with the state preparation, we use a half-wave plate denoted as HWP$_\alpha$ to prepare the initial state of photons in the state $ \ket{\psi_i}=(a\ket{H}+b\ket{V})\ket{r}$, where $a$ and $b$ are real coefficients, and we assume  $\phi_i=\phi_f=0$. Here, $\ket{H}$ and $\ket{V}$ are the horizontal and vertical polarizations, respectively, $\ket{r}$ is the initial path degree of freedom and $|a|^2+|b|^2=1$.
The photons are sent to a set of beam displacers (BD), which deviate the photons spatially depending on the  polarization. The direction of this deflection is related to the orientation of the optical axis of each BD. For instance, BD1 transmits the $\ket{V}$ photons, maintaining  the path $\ket{r}$,  while deviates horizontally  (parallel to the optical table) the $\ket{H}$ photons. Thus, the output state of BD1 is $ \ket{\psi_1}= a\ket{H}\ket{l}+b\ket{V}\ket{r}$, where $\ket{l}$ is the path of the deviated photons. This path degree of freedom (DoF) encodes the system  by interpreting $\ket{l}$ ($\ket{r}$) as the ground state $\ket{g}$ (excited state $\ket{e}$). After,  the photons   pass through a half-waveplate (HWP$_{\phi}$), which rotates the polarization by an angle $\phi$, and   BD2 that deviates the photons vertically (perpendicular to the optical table). This beam-displacer populates the upper layer of the interferometer \cite{aguilar2022two}. We utilized $\ket{u}$ and $\ket{d}$ to refer to the upward and downward paths, which codify the reservoir energy states $\ket{\mathcal{E}_e}$ and $\ket{\mathcal{E}_g}$, respectively.  It is worth noting that, after ignoring the polarization, the state of the photons after BD2 is the initial product state required in the forward process [see Eq.~(4) of the main text]. Therefore, after BD2  the initial system-environment state is
\begin{multline}
\hat{\rho}_{S\mathcal{E}}=\hat{\rho}_S\otimes\hat{\rho}_{\mathcal{E}}\\
=|a|^2\sin^2\phi\ket{ld}\bra{ld}+ab^{*}\sin^2\phi\ket{ld}\bra{rd}+a^{*}b\sin^2\phi\ket{rd}\bra{ld}\\
+|b|^2\sin^2\phi\ket{rd}\bra{rd}+|a|^2\cos^2\phi\ket{lu}\bra{lu}+ab^{*}\cos^2\phi\ket{lu}\bra{ru}\\
+a^{*}b\cos^2\phi\ket{ru}\bra{lu}+|b|^2\cos^2\phi\ket{ru}\bra{ru}.
\end{multline}
where $\rho_i$ and  $\rho_{th}$ are mapped into the states
\begin{eqnarray}
\hat{\rho}_S&=&|a|^2\ket{l}\bra{l}+|b|^2\ket{r}\bra{r}+ab(\ket{r}\bra{l}+\ket{l}\bra{r}),\\
 \hat{\rho}_\mathcal{E}&=&\sin^2(\phi)\ket{d}\bra{d}+\cos^2(\phi)\ket{u}\bra{u},
\end{eqnarray} 
 respectively. The comparison between these operators and the theoretical description in Eqs.~(2) and~(3) allows the following correspondence 
\begin{eqnarray}
\cos{\frac{\theta_i}{2}}=a &,& \sin{\frac{\theta_i}{2}}=b,\\
    w_g= \sin^2(\phi) &,& w_e= \cos^2(\phi).
    \end{eqnarray}
\section*{b) ~ States evolution}

In the second stage of the experiment we make the system and reservoir interact via the unitary operation $\hat{U}$ in Eq.~(5). This is implemented with  BD3, BD4 and a set of HWPs. The dissipation rate, determined by the damping parameter $p\in[0,1]$ \cite{almeida2007environment,salles2008experimental,aguilar2014experimental,aguilar2014flow}, is controlled with the HWP$_{\theta}$, by using the parametrization $p=\cos^2{\theta/2}$. Optical path compensation is achieved by inserting two HWP$_0$ at zero. Note that all parameters are experimentally controlled by rotating various wave plates. The evolved state of the system and environment $(\hat{\rho}'_{S\mathcal{E}})$ in Eq. (4), is obtained at the output of BD4, and is given as 
\begin{multline}
\hat{\rho}'_{S\mathcal{E}}=(a \sin\phi\ket{ld}+b\sin\phi\cos\theta\ket{lu}+b\sin\phi\sin\theta\ket{rd}).\\(a^{*}\sin\phi\bra{ld}+b^{*}\sin\phi\cos\theta\bra{lu}+b^{*}\sin\phi\sin\theta\bra{rd}) \\   
+(a\cos\phi\sin\theta\ket{lu}+a\cos\phi\cos\theta\ket{rd}+b\cos\phi\ket{ru}).\\(a^{*}\cos\phi\sin\theta\bra{lu}+a^{*}\cos\phi\cos\theta\bra{rd}+b^{*}\cos\phi\bra{ru}).
\end{multline}
Through partial trace, the reduced density matrices of the evolved system ($\hat{\rho}'_S$) and environment ($\hat{\rho}'_E$) are
\begin{multline}
 \hat{\rho}'_S=Tr_{\mathcal{E}} \left[\hat{\rho}'_{S\mathcal{E}}\right]=\\(a^2\sin^2\phi+b^2\sin^2\phi\cos^2\theta+a^2\cos^2\phi\sin^2\theta)\ket{l}\bra{l}\\
 +(b^2\sin^2\phi\sin^2\theta+a^2\cos^2\phi\cos^2\theta+b^2\cos^2\phi)\ket{r}\bra{r}\\
 +ab\sin\theta(\ket{l}\bra{r}+\ket{r}\bra{l}),
 \label{evolved_system}
\end{multline}

\begin{multline}
\hat{\rho}'_{\mathcal{E}}=Tr_S \left[\hat{\rho}'_{S\mathcal{E}}\right]=\\
(a^2\sin^2\phi+b^2\sin^2\phi\sin^2\theta+a^2\cos^2\phi\cos^2\theta)\ket{d}\bra{d}\\
+(b^2\sin^2\phi\cos^2\theta+a^2\cos^2\phi\sin^2\theta+b^2\cos^2\phi)\ket{u}\bra{u}\\
+ab\cos\theta(\ket{d}\bra{u}+\ket{u}\bra{d}). \label{evolved_envir}
\end{multline}
Here the terms with "ab" in the density states of both the system and the environment are associated with coherence, and their control is governed by the HWP$_{\theta}$. When the parameter ``$p$" is equal to 1, the system completely loses its coherence to the environment.

\section*{c) ~ Projective measurements}
Finally, projective measurements are carried out by the final pair of BDs (BD5 and BD6), in conjunction with a couple of  HWPs. A quarter-wave plate (QWP) is also inserted to compensate for undesired phases that appear in the propagation of the photons inside the interferometer.  
By adjusting the angle of the HWP, the experimental setup allows  to perform  projections onto any linear polarization of the system and onto each state of the energy eigenbasis of the reservoir. The corresponding forward $ P_{F}(\psi_{f}|\psi_{i})$ and backward probabilities $P_{B}(\bar{\psi}_{f}|\bar{\psi}_{i})$ for the three different cases can be written as: 
\begin{eqnarray}
P_{F}(V|H)&=&Tr[\hat{\rho}'_{S}\ket{V}\bra{V}]=\cos^2\phi\cos^2\theta,\\
P_{B}(H|V)&=&Tr[\hat{\rho}'_{S}\ket{H}\bra{H}]=\cos^2\theta\sin^2\phi,\\
P_{F}(V|D)&=&Tr[\hat{\rho}'_{S}\ket{V}\bra{V}]=\frac{1}{2}((1+\cos^2\theta)\cos^2\phi\nonumber\\
&+&\sin^2\theta\sin^2\phi),\\
P_{B}(D|V)&=&Tr[\hat{\rho}'_{S}\ket{D}\bra{D}]=\frac{1}{2}.
\end{eqnarray}

Finally for the third case, namely $ \ket{\psi_{i}}=1/\sqrt{2} (\ket{g}+\ket{e})$ and  $ \psi_{f}=1/2( \ket{g}+ \sqrt{3} \ket{e})$, the forward and backward probabilities can be written as:
\begin{eqnarray}
P_{F}(\psi_{f}|\psi_{i})&=&Tr[\hat{\rho}'_{S}\ket{\psi_{f}}\bra{\psi_{f}}] \nonumber \\
&=&\frac{1}{16} (8 + \cos\left(2(\theta - \phi)\right) + 2\cos\left(2\phi\right) \nonumber \\ 
&+& \cos\left(2(\theta + \phi)\right) + 4\sqrt{3}\sin(\theta)),\\
P_{B}(\bar{\psi}_{f}|\bar{\psi}_{i})&=&Tr[\hat{\rho}'_{S}\ket{\bar{\psi}_{f}}\bra{\bar{\psi}_{f}}] \nonumber\\
&=&\frac{1}{4} \left(2 + \sqrt{3}\sin(\theta)\right).
\end{eqnarray}

\raggedright
\normalem


%

\end{document}